\documentclass[pre]{revtex4}
\usepackage{graphicx}

\newcommand{\cmc} {\varphi_{\rm cmc}}
\newcommand{\eg} {{e.g., }}
\newcommand{\half} {\frac{1}{2}}
\newcommand{\ie} {{i.e., }}
\newcommand{\kT} {k_{\rm B}T}
\newcommand{\meom} {$m$-EO$_x$-$m$ }
\newcommand{\msm} {$m$-$s$-$m$ }
\newcommand{\rmB} {{\rm B}}
\newcommand{\rmc} {{\rm c}}
\newcommand{\rmd} {{\rm d}}
\newcommand{\rmD} {{\rm D}}
\newcommand{\rme} {{\rm e}}

\newcommand{\rmm} {{\rm m}}
\newcommand{\rms} {{\rm s}}
\newcommand{\rmS} {{\rm S}}

\begin{document}

\title
{Models of Gemini Surfactants}
\thanks{To be published in
{\it Gemini Surfactants: Interfacial \& Solution Phase Behavior},
R.\ Zana and J.\ Xia (eds.), Marcel Dekker, New York (2003).}

\author{Haim Diamant$^1$}
\author{David Andelman$^2$}

\affiliation{$^1$School of Chemistry and $^2$School of Physics
\& Astronomy, Beverly and Raymond Sackler Faculty of Exact Sciences, 
Tel Aviv University,
Tel Aviv 69978, Israel}

\maketitle

\section{Introduction}

Gemini surfactants are composed of two monomeric surfactant
molecules linked by a spacer chain. They constitute a new class
of amphiphilic molecules having its own distinct behavior. Since
their first systematic studies over a decade ago, gemini
surfactants have been the subject of intensive research (see Ref.\
\cite{Zana_review} and references therein).  Research has been
motivated by the advantages of gemini surfactants over regular
ones with respect to various applications, \eg their increased
surface activity, lower critical micelle concentration (cmc), and
useful viscoelastic properties such as effective thickening.

Besides their importance for applications, the behavior of gemini
surfactants is qualitatively different in several respects from
that of regular surfactants, posing challenges to current
theories of surfactant self-assembly. The main puzzles raised by
gemini surfactants can be summarized as follows
\cite{Zana_review}.
\begin{itemize}

\item {\it Surface behavior}. The area per molecule in a saturated
monolayer at the water--air interface, made of gemini surfactants
with polymethylene spacers (\msm surfactants, where $s$ is the
spacer length and $m$ the tail length in hydrocarbon groups), has
a non-monotonous dependence on $s$ \cite{Alami,Perez}. For
example, for tail length of $m=12$ the molecular area at the
water--air interface is found to increase with $s$ for short
spacers, reach a maximum at about $s\simeq$ 10--12, and then
decrease for longer spacers. This decrease in the
specific area for the \msm surfactants is somewhat unexpected given
the fact that the molecule becomes bigger as $s$ increases. One
would naively expect a monotonous increase in the molecular area
as indeed is observed for another class of gemini surfactants
having a poly(ethylene oxide) spacer (\meom surfactants)
\cite{Dreja99,Zhu}.

\item {\it Micellization point}. The critical micelle
concentration (cmc) of gemini surfactants is typically one to two
orders of magnitude lower than that of the corresponding monomeric
surfactants having the same head and tail groups \cite{Zana91}. For
regular (monomeric) surfactants the cmc decreases monotonously with
the number of hydrocarbon groups because of increased
molecular hydrophobicity. In the case of \msm gemini surfactants, by
contrast, the dependence of the cmc on the spacer length $s$ is
non-monotonous with a maximum at about $s\simeq$ 4--6
\cite{Devinsky,Zana91,De}.  Similarly, the Krafft
temperature exhibits a minimum \cite{Zana02} and the micellization
enthalpy a maximum \cite{Grosmaire} at about the same $s$ value.

\item {\it Aggregate shape}. As certain parameters, such as the
relative size of the head and tail groups or the salt concentration,
are progressively changed, regular surfactants change their aggregate
morphology in the direction of decreasing curvature, \eg from
spherical micelles to cylindrical micelles to bilayer vesicles
\cite{Israelachvili,Israelachvili_book}.  However, when the
polymethylene spacer length in \msm gemini surfactants is increased, a
different sequence of shapes is observed, for instance, from
cylindrical micelles to spherical micelles to vesicles for the
12-$s$-12 surfactants \cite{Zana93,Danino}. Moreover, gemini
surfactants with short spacers exhibit uncommon aggregate morphologies
in the form of branched cylindrical micelles and ring micelles
\cite{Ann}.

\item {\it Phase behavior}.
The spacer length in \msm gemini surfactants has an unusual effect
also on the phase behavior of binary surfactant--water mixtures.  For
geminis with tail length $m=12$, for instance, the phase-diagram
region corresponding to hexagonal and lamellar phases is found to
shrink with increasing $s$, disappear for $s=10$--$12$, and re-appear
for $s=16$ \cite{Zana93b}. In ternary systems of water--oil--\msm
surfactant, the size of the microemulsion (single-phase) region in
the phase diagram has a non-monotonous dependence on $s$ with a
maximum at $s\simeq 10$ \cite{Dreja98a}.

\item {\it Dynamics}.
Dilute micellar solutions of gemini surfactants with short spacers
have unusual rheological properties, such as
pronounced increase in viscosity upon increase of surfactant
volume fraction \cite{Kern,In} and shear-thickening
\cite{Schmitt,Oelschlaeger}.

\end{itemize}

In view of the amount of experimental work and its unusual findings,
the number of theoretical studies devoted to gemini surfactants has
been surprisingly small. In this chapter we have, therefore, two
aims. The first is to review the current state of theoretical models
of gemini surfactants.  The second, perhaps more important aim, is to
indicate the considerable gaps in our knowledge and the key open
questions awaiting theoretical work. In Sec.\ \ref{sec_regular} we set
the stage by reviewing several theoretical models of surfactant
self-assembly. This will facilitate the discussion in Sec.\
\ref{sec_gemini} of the gemini surfactant models, which can be
viewed as extensions to previous models of regular
surfactants. Finally, in Sec.\ \ref{sec_conclusion} we conclude and
summarize the open questions.

\section{Models of surfactant self-assembly}
\label{sec_regular}

In this section we review several theoretical models pertaining to the
self-assembly of soluble surfactants. This is not meant to be an
exhaustive review of self-assembly theory but merely to provide the
appropriate background for the models of gemini surfactants discussed
in Sec.\ \ref{sec_gemini}.

\subsection{Surface behavior}
\label{sec_surface_reg}

Let us start by considering an aqueous surfactant solution below the
cmc.  The soluble surfactant molecules self-assemble into a condensed
layer at the water--air interface, referred to as a {\it Gibbs
monolayer} (to be distinguished from {\it Langmuir monolayers} that
form when insoluble surfactants are spread on the water--air
interface) \cite{Adamson}.  Since the surfactant is water-soluble,
this layer exchanges molecules with the bulk solution and a nonuniform
concentration profile forms.  Typical surfactants have strong surface
activity, \ie the energy gained by a molecule when it migrates to the
surface is much larger than the thermal energy $\kT$. As a result, the
concentration profile drops sharply to its bulk value within a
molecular distance from the surface (hence the term {\it monolayer}).

The number of molecules participating in a Gibbs monolayer per unit
area, the surface excess $\Gamma$, is obtained from integrating the
excess concentration (with respect to the bulk one) over the entire
solution.  Such a monolayer can be regarded as a separate sub-system
at thermodynamic equilibrium and in contact with a large reservoir of
molecules having temperature $T$ and chemical potential $\mu$. From
the excess free energy per unit area of this system, $\gamma(T,\mu)$,
which is by definition the surface tension of the solution, we get the
number of molecules per unit area:
\begin{equation}
  \Gamma = -\left( \frac{\partial\gamma}{\partial\mu} \right)_T.
\end{equation}
This is referred to as the Gibbs equation \cite{Adamson}.  For dilute
solutions $\mu\propto\kT\ln C$, where $C$ is the bulk surfactant
concentration. (The constant of proportionality is one for nonionic
surfactants and ionic ones at high salt concentration; it has a higher value
for salt-free ionic surfactant solutions, where strong correlations
between the different ions lead to non-ideal activity coefficients
\cite{Adamson}.) Hence,
\begin{equation}
  \Gamma \propto -\frac{1}{\kT}
  \left( \frac{\partial\gamma}{\partial\ln C} \right)_T.
\end{equation}
Because of the high surface activity of surfactant molecules, leading
to a sharp concentration profile at the water--air interface,
$\Gamma^{-1}$ is commonly interpreted as the average surface area per
molecule, $a$.  The second consequence of the high surface
activity is that, already for $C$ much smaller than the cmc, the
monolayer becomes saturated, \ie $\Gamma$ stops increasing with
$C$. Experimentally, the curve describing the change in $\gamma$ as a
function of $\ln C$ becomes a straight line with a negative slope
proportional to $-\Gamma$.

We now wish to find a simple estimate for the energy of lateral
interaction between molecules in such a saturated monolayer (repeating
a well-known result of Ref.~\cite{Israelachvili}).  Saturation implies
that the molecules are packed in an energetically optimal density,
such that there is no gain in adding or removing molecules. This
optimum arises from a competition between attractive and repulsive
interactions. The attractive interaction tries to decrease the area
per molecule, and we can phenomenologically write its energy per
molecule as proportional to $a$, $\gamma_1a$, where the
proportionality constant $\gamma_1$ has units of energy per unit
area.  Since the attraction comes mainly from the desire of the
hydrocarbon tails to minimize their contact with water, $\gamma_1$
should be roughly equal to the hydrocarbon--water interfacial tension
($\gamma_1\simeq 50$\,mN/m).  The repulsive interaction, on the other
hand, tries to increase $a$ and, at the same phenomenological
level, we can write its energy per molecule as inversely proportional
to $a$, $\alpha/a$, where
$\alpha$ is a positive constant. Minimizing the sum of these two
contributions we get for the interaction energy per molecule
$u=\gamma_1(a-a_0)^2/a +
\mbox{const}$, where $a_0=(\alpha/\gamma_1)^{1/2}$ is the optimal
molecular area. Expanding around $a=a_0$ to second order and
omitting the constant term, we obtain
\begin{equation}
  u(a) \simeq \frac{\gamma_1}{a_0}(a-a_0)^2.
\end{equation}
This is merely a harmonic approximation for the interaction energy
associated with small deviations from optimal packing.

We can slightly modify this result to obtain a similar harmonic
estimate for the energy $u_2$ of effective interaction between {\it
two} neighboring molecules residing in the saturated monolayer. (This
will be useful later on when we add the spacer to form gemini
surfactants.) We need to divide $u$ by half the number of inplane
neighbors, $q/2$, and express $a$ in terms of the average
intermolecular distance $r$, $a=\eta r^2$, where $\eta$ is a prefactor
of order unity (\eg for hexagonal packing $q=6$ and
$\eta=\sqrt{3}/2\simeq 0.9$). Assuming again that $r$ is close to its
optimal value $r_0$, we get
\begin{equation}
  u_2(r) \simeq \half k_0 (r-r_0)^2, \ \ \ \
  k_0 = \frac{16\eta\gamma_1}{q}.
\label{ki}
\end{equation}
This expression replaces the actual surfactant--surfactant interaction
in the monolayer, including effects of other nearby surfactants, with
an ``effective spring'' of equilibrium length $r_0$ and spring constant
$k_0$.  For hexagonal packing we get the reasonable value $k_0\simeq
120$\,mN/m $\simeq 0.3$ $\kT$/\AA$^2$. Note that, because of the 
saturation condition, the expression for $k_0$ is insensitive to molecular
details. In turn, those details will affect the properties of the saturation
state itself, \eg the value of $a_0$ or $r_0$.

\subsection{Micelles}
\label{sec_micelle_reg}

As the solution concentration is increased beyond the cmc the
surfactant molecules start to form aggregates. Unlike simple solute
molecules (\eg alkanes), which undergo macroscopic phase separation
upon increasing concentration or changing temperature, surfactants
form micelles at the mesoscopic scale.  The challenges posed to
theories of surfactant self-assembly are to predict the micellization
point as a function of concentration (\ie the cmc, hereafter referred
to by the corresponding volume fraction $\cmc$) and temperature
($T_\rmm$), as well as the micelle shape and size. The main
complications come from the fact that micellization is {\it not} a
macroscopic phase transition --- the aggregate sizes are finite and
polydisperse --- and thus the well-developed theoretical framework of
phase transitions does not strictly apply.

From a thermodynamic point of view, the difference between surfactant
micellization and phase separation lies in the following observation
\cite{Israelachvili_book}. For alkanes solubilized in water, for
example, the (Gibbs) free energy per molecule in an aggregate of size
$N$, $g_N$, is a monotonously decreasing function of $N$ --- for
$N\rightarrow\infty$ $g_N$ tends to the free energy per molecule in
the bulk alkane phase, $g_\infty$, while for smaller $N$
$g_N>g_\infty$ due to unfavorable surface terms of the finite cluster.
As a result, there is a critical concentration (or critical
temperature) at which the favorable size changes discontinuously from
monomers solubilized in water ($N=1$) to a macroscopic phase of bulk
alkane ($N\rightarrow\infty$).  The first-order phase transition
point is reached when the chemical potential of monomers exceeds
$\mbox{min}\{g_N\}=g_\infty$.  In a dilute solution this implies that
$\varphi_\rmc=\rme^{g_\infty/(\kT)}$, $\kT_\rmc=g_\infty/\ln\varphi$.
(We have set the free energy of the $N=1$ state as the reference,
$g_1=0$.) In the case of surfactants, by contrast, $g_N$ has a minimum
at a finite aggregate size $N^*$.  As a result, when the chemical
potential exceeds $g_{N^*}$, a large population of aggregates appears,
whose sizes are distributed around $N^*$. Hence, the micellization
point can be estimated as
\begin{equation}
  \cmc = \rme^{g_{N^*}/(\kT)},\ \ \ \
  \kT_\rmm = g_{N^*}/\ln\varphi.
\label{micellization}
\end{equation}

The remaining task is to obtain a theoretical expression for
$g_N^{(\rmS)}$, the free energy per molecule in aggregates of size $N$
and shape S. Since we expect this function to have a minimum at a
finite yet large $N$ (say, $N^*\sim 10^2$), the importance of
many-body interactions is inevitable, and obtaining $g_N^{(\rmS)}$
from rigorous statistical mechanics is a formidable
task. Consequently, analytical models have relied on phenomenological
approaches, trying to account for various competing contributions to
the free energy while assuming a certain geometry for the aggregate
\cite{Israelachvili_book,Israelachvili,Nagarajan1}.  From the
minimum of $g_N^{(\rmS)}$ with respect to $N$ and various
possible shapes S one can obtain the 
aggregate shape, aggregation number, and micellization point,
using Eq.\ (\ref{micellization}).

\begin{figure}[tbh]
\vspace{.5cm}
\centerline{\resizebox{0.17\textwidth}{!}
{\includegraphics{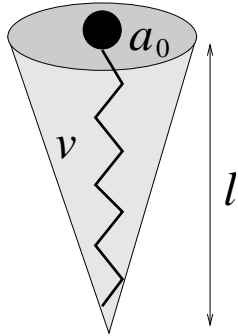}}} 
\caption[]
{Packing constraints on a surfactant in an aggregate. Each head group
occupies an optimal area $a_0$ on the aggregate surface; the tail
chain occupies a volume $v$ and cannot stretch beyond length
$l$. These constraints define the packing parameter, $P=v/(a_0 l)$,
which suggests the possible aggregate shape.}
\label{fig_packing_parameter}
\end{figure}

In the simplest picture \cite{Israelachvili_book,Israelachvili} rough
estimates for the minimum of $g_N^{(\rmS)}$ can be obtained by imposing
geometrical constraints that arise from the incompressibility of the
micellar hydrocarbon core (see Fig.\
\ref{fig_packing_parameter}).  These constraints lead to a finite 
aggregation number $N^*$, as required. The hydrocarbon tail chains
cannot extend beyond a certain length $l$, and each tail must occupy a
certain volume $v$. (Both $l$ and $v$ are known to have a simple
linear dependence on the number of hydrocarbon groups in the tail
chain \cite{Tanford}.) In addition, as in the case of the Gibbs
monolayer of Sec.\ \ref{sec_surface_reg}, surfactant molecules in the
favorable aggregate state attain a certain optimal area per molecule
$a_0$. Then, due to translational entropy, the favorable
aggregates would be the smallest ones that satisfy all
constraints. The constraints define a dimensionless {\it packing
parameter} (Fig.\ \ref{fig_packing_parameter}),
\begin{equation}
  P = \frac{v}{a_0 l}.
\label{packing}
\end{equation}
If $P<1/3$ the constraints can be satisfied by spherical micelles,
which will be the smallest and hence the most favorable ones; when
$1/3<P<1/2$ the micelles will have to be of elongated or cylindrical shapes;
for $1/2<P<1$ planar shapes will form; and for $P>1$ the morphology
must be inverted.

Once the shape is determined we can find the
maximum allowed aggregation number. For example, for spherical micelles
$N< 4\pi l^2/a_0 =(4\pi/3)l^3/v$, and we get
\begin{equation}
  N < 36\pi \frac{v^2}{a_0^3}.
\end{equation}
We see how competing interactions between the molecules (giving
rise to $a_0$) together with the incompressibility of the micellar
core lead to finite micelles.  Since the tail chains usually
should not stretch to their full extent, the actual aggregation number
will be smaller than this upper bound.

Yet, these geometrical arguments cannot provide us with theoretical
predictions as to the optimal molecular area $a_0$ itself or the
aggregation free energy $g_{N^*}^{(\rmS)}$, as well as their
dependence on parameters such as temperature or salt concentration. In
order to get such information and subsequently predict the
micellization point, micelle shape and size, one needs a more detailed
theory.

\subsubsection{Phenomenological models}

There have been attempts to analytically account for the various
competing contributions to the free energy per molecule
$g_N^{(\rmS)}$ (\eg Ref.\ \cite{Nagarajan1}). The advantage of this
approach is that, once we have an expression for the free energy, we
can easily change parameters and gain insight into the role of various
contributions.  On the other hand, such models essentially attempt to
push the limits of the phenomenological approach toward a detailed
molecular description. They usually entail uncontrolled approximations
and parameters whose accurate values are often hard to obtain.  As an
example, which will serve us in Sec.\ \ref{sec_gemini}, we give here
an analysis along the lines of (yet not identical to) Ref.\
\cite{Nagarajan1}.

Five major contributions to the aggregation free energy (per
surfactant molecule on the aggregate) can be considered
\cite{Nagarajan1},
\begin{equation}
  g_N^{(\rmS)}(a) = g_{\rm hc} + g_{\rm int}(a) +
  g_{\rm es}(a,R) + g_{\rm st}(a) + g_{\rm el}(R),
\end{equation}
where $a$ is the area per molecule on the aggregate surface and
$R$ the aggregate size (radius or width). Note that $R$ is not an
independent variable but is related to $N$ and $a$ via the
aggregate geometry S, \eg for spherical micelles, $Na=4\pi
R^2$.

(i) The driving force for aggregation is the hydrophobic effect, \ie
the free energy per surfactant molecule $g_{\rm hc}$ gained by
shielding the hydrocarbon groups from water \cite{Tanford}.  This
contribution to $g_N^{(\rmS)}$ is negative and, to a good
approximation, independent of $N$ and the aggregate geometry S.
Namely, its contribution to the {\it entire} aggregate free energy is
linear in $N$ and tends to increase the aggregate size.  The
hydrophobic term $g_{\rm hc}$ depends linearly on the number of
hydrocarbon groups in the surfactant, with a reduction of roughly
$\kT$ per hydrocarbon group \cite{Israelachvili_book}. That is why,
for regular surfactants, the cmc decreases exponentially with the
number of hydrocarbon groups in the molecule and is reduced by a
factor of roughly 2--3 per each additional hydrocarbon group.

(ii) The hydrophobic gain is corrected by an interfacial contribution
$g_{\rm int}$ due to the unfavorable contact between the hydrocarbon
core and water,
\begin{equation}
  g_{\rm int}(a) = \gamma_1 (a - a_{\rm min}),
\end{equation}
where $\gamma_1$ is the interfacial tension of the core--water
interface (roughly equal to the hydrocarbon--water interfacial
tension), and $a_{\rm min}$ is the minimum area per molecule, \ie
the interfacial area occupied by a head group. This contribution
evidently acts to reduce the area per molecule.

(iii) If the surfactant head groups are charged, there is
electrostatic repulsion between them, acting to increase
$a$. Within the Poisson--Boltzmann theory this electrostatic
contribution is given by
\cite{Ninham}:
\begin{equation}
  g_{\rm es}(a,R) = 2\kT \left\{ \ln[\beta+(1+\beta^2)^{1/2}] -
  [(1+\beta^2)^{1/2}-1]/\beta - \frac{2c\lambda_\rmD}{\beta} \ln
  \left[ \half + \half (1+\beta^2)^{1/2} \right]\right\},
\end{equation}
where $\beta=4\pi l_\rmB \lambda_\rmD/a$ is a dimensionless
charging parameter depending on two other lengths, the Debye screening
length $\lambda_\rmD$ and the Bjerrum length $l_\rmB$. The Debye
screening length in the solution is $\lambda_\rmD=(8\pi l_\rmB c_{\rm
salt})^{-1/2}$, where $c_{\rm salt}$ is the added salt concentration,
taken here to be monovalent, and $l_\rmB = e^2/(\varepsilon\kT)$
is about 7 \AA~for aqueous solution with dielectric constant
$\varepsilon=80$ at room temperature. (For simplicity, a monovalent
head group has been assumed.) Finally, $c$ is the mean curvature of
the aggregate (\eg $1/R$ for spherical micelles).

(iv) There is also steric repulsion between head groups. From the
(non-ideal) entropy of mixing per molecule we get for this
contribution
\begin{equation}
  g_{\rm st}(a) = \kT \left[ \ln(a_{\rm min}/a) +
  (a/a_{\rm min}-1) \ln(1-a_{\rm min}/a) \right].
\end{equation}

(v) The last contribution to the free energy is associated with
the tail packing in the hydrophobic core, \ie deviations
of the hydrocarbon tail chains from their relaxed length $l_0$,
\begin{equation}
  g_{\rm el}(R) = \half k' (R-l_0)^2.
\end{equation}
The ``elastic constant'' $k'$ depends on the chain statistics, as well as
the packing parameter (\ie aggregate shape) \cite{Nagarajan1}.

The equilibrium aggregation number $N$ and specific area $a$ (and
hence also aggregate size $R$) for a given shape S and surfactant
chemical potential $\mu$ are then determined by the equations
\begin{equation}
  g_N^{(\rmS)}(a) = \mu, \ \ \ \
  \frac{\partial g_N^{(\rmS)}(a)}{\partial a} = 0.
\end{equation}
Comparing the minimum value of $g_N^{(\rmS)}$ for various shapes S,
one also obtains the equilibrium aggregate morphology. As long as
$\mu<\kT\ln\cmc$, these equations will have no solution, and the
monomeric state ($N=1$) of single surfactant molecules solubilized in
water is the stable one. As the chemical potential increases, we reach
the micellization condition given by Eq.\ (\ref{micellization}), where
the average micelle size at the cmc, $N^*$, can be calculated now from
the expression of $g_N^{(\rmS)}$ at its minimum.

\subsubsection{Computer simulations}

Another route to overcome the complexity of treating surfactant
micellization is to use computer simulations. This approach can be
divided into two categories: statistical-mechanical models using Monte
Carlo (MC) simulations, and Molecular Dynamics (MD) simulations.

\begin{figure}[tbh]
\vspace{.5cm}
\centerline{\resizebox{0.27\textwidth}{!}
{\includegraphics{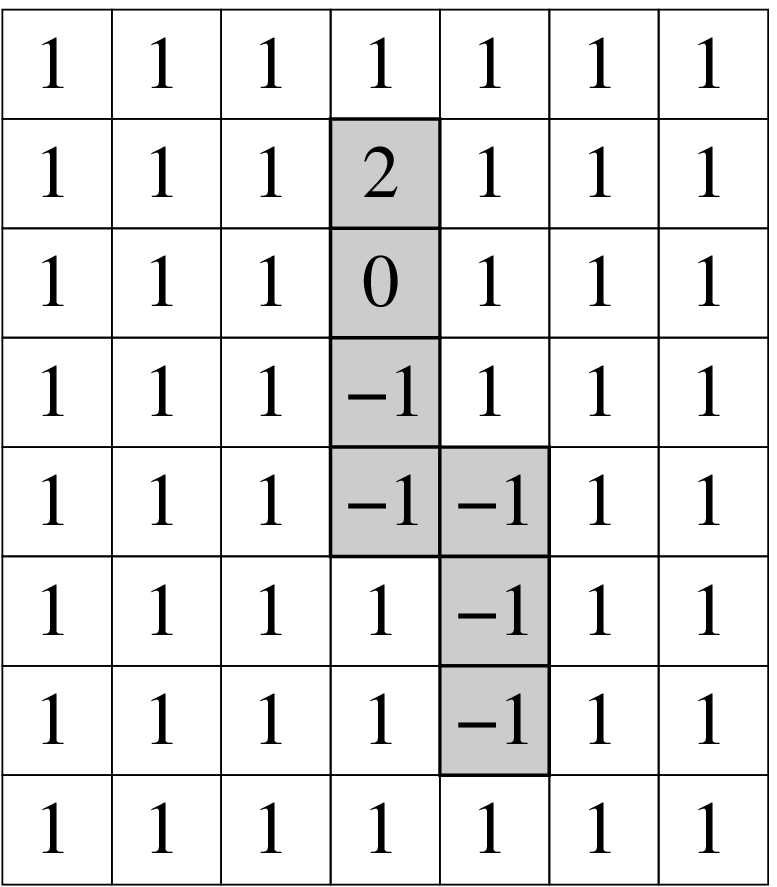}}}
\caption[]
{Schematic representation of a surfactant molecule (gray)
solubilized in water (white) in a lattice spin model
\cite{Stauffer94a}. Each water
molecule and surfactant group are represented by a spin variable on a
lattice site. Water molecules have spin $+1$, head groups $+2$, and
tail groups $-1$. In between the head and the tail there is a
``neutral" group of spin $0$. The various particles interact via
nearest-neighbor ``ferromagnetic'' couplings favoring spins of the
same sign, except for the head--head interaction which is
``antiferromagnetic'' disfavoring neighboring heads. The chain
connectivity and the overall number of chains are preserved during the
MC simulation.}
\label{fig_Stauffer}
\end{figure}

Following Widom's statistical-mechanical model of microemulsions
\cite{Widom}, a host of lattice models were presented
for treating surfactant self-assembly (see, \eg Refs.\
\cite{Larson1,Larson2,Stauffer93,Stauffer94a,Stauffer94b}).
These molecular ``toy models'' represent the water molecules and
various groups in the surfactant as Ising spins on a discrete
lattice. The various interactions between the groups are represented
by ferromagnetic or antiferromagnetic couplings between {\it
nearest-neighbor} spins (see Fig.\ \ref{fig_Stauffer}).  Evidently,
this is a very crude description of surfactant solutions and is not
expected to yield quantitative predictions. Another difficulty is
attaining thermodynamic equilibrium in simulations of these
self-assembling systems, which contain slowly relaxing, large
aggregates.  Such models, however, have been shown to correctly
reproduce various qualitative features of amphiphilic systems, \eg
aggregate formation, aggregate shape, and the overall structure of
phase diagrams.  The main advantage of this statistical-mechanical
approach is that, by tuning a small number of parameters, one can get
from the MC simulations insight into molecular mechanisms that
determine the overall system behavior.  Here we briefly present a
model similar to that of Ref.\
\cite{Stauffer94a}. It will serve us when we discuss gemini
surfactants in Sec.\ \ref{sec_micelle_gemini}.

In the lattice model each water molecule is assigned a spin $\sigma=+1$, a
hydrocarbon group in the tail has spin $\sigma=-1$, and the head group
has spin $\sigma=+2$ (see Fig.\ \ref{fig_Stauffer}). In between the
hydrophilic head group and the hydrophobic tail there is a `neutral'
group of spin $\sigma=0$. All the couplings are ferromagnetic
(favoring neighboring spins of the same sign), except for the
head--head coupling which is antiferromagnetic, mimicking head--head
repulsion due to screened electrostatics. The energy of the system can
be written as
\begin{equation}
  E = -J \sum_{\langle ij\rangle} \sigma_i\sigma_j
  (1 - 2\delta_{\sigma_i,2}\delta_{\sigma_j,2}),
\label{Ising}
\end{equation}
where $J>0$ is the coupling strength, $\langle ij\rangle$ denotes
summation only over nearest-neighbor pairs of the lattice, and
$\delta_{i,j}=1$ when $i=j$ and zero when $i\neq j$ is the Kronecker
delta function.  Two neighboring water molecules attract each other
with energy $-J$ because then $\sigma_i\sigma_j=1$.  The same applies
to two tail groups, while a water molecule and a tail group repel with
energy $+J$. The extra factor of $(1 -
2\delta_{\sigma_i,2}\delta_{\sigma_j,2})$ in Eq.\ (\ref{Ising}) is
unity for all cases except when $i$ and $j$ are two heads with
$\sigma=+2$, yielding a repulsion of $+4J$ between two head groups. 
Finally, a head group and a water molecule attract with energy
$-2J$, and a head group and a tail group repel with energy
$+2J$. These couplings apply whether the two neighboring lattice sites
belong to the same molecule or not. In addition, the groups belonging
to the same surfactant molecule are kept linked throughout the
simulation. Thus, the essential features of hydrophobicity,
hydrophilicity, molecular connectivity, and (screened) electrostatic
repulsion are all accounted for using the single parameter $J$. Other
parameters are the length of the tail group and total number of
surfactant molecules in the system.

The MC simulation starts from a certain configuration of surfactant
molecules in water.  At each iteration the various groups of the
surfactants are moved while maintaining the connectivity of the
molecules, their total number, and the total number of water molecules
(\eg using a ``slithering snake'' scheme \cite{Stauffer94a}).  The
energetic cost of the move is calculated using Eq.\ (\ref{Ising}), and
the MC step is accepted or rejected according to a Metropolis
criterion, ensuring convergence toward equilibrium.

This simple scheme can reproduce much of the richness of surfactant
self-assembly, including the formation of monolayers, micelles and
bilayers, the dependence of the cmc on tail length, transitions
between various aggregate shapes, etc. On the other hand, such models
can merely indicate general trends and not detailed information. For
example, the correspondence between the MC spin variable representing
sub-groups of a surfactant molecule and the actual chemical groups is
not well defined and remains ambiguous to some extent.

Another class of numerical studies that have been used to explore
surfactant self-assembly are Molecular Dynamics (MD)
simulations. These models range in detail from coarse-grained
``bead--spring'' representations of the molecules (\eg
\cite{Smit90,Smit93,Palmer,Fodi,Karaborni,Maiti02}) to atomistic
descriptions (\eg \cite{Watanabe,Shelley,Mackerell,Maillet,Oda}).  The
advantage of the MD approach, as compared to phenomenological theories
and spin models, is that the description of the system on the
molecular scale is less artificial.  The disadvantages are the limited
spatial and temporal extent of the simulations, entailing
equilibration problems, and sometimes also a large number of required
parameters.  A typical all-atom MD simulation of an aqueous surfactant
system may contain about a hundred surfactant molecules along with few
thousands water molecules, and the dynamics can be run for a few
nanoseconds (\eg Ref.\ \cite{Oda}).  A coarse-grained simulation
allows a significant increase of these numbers at the expense of
molecular detail (see, \eg Ref.\ \cite{Maiti02}). Here we outline a
coarse-grained approach to surfactant micellization, as presented in
Ref.\ \cite{Smit90}, which was later extended to gemini surfactants
\cite{Karaborni}.

The MD model of Ref.\ \cite{Smit90} contains only two types of
``particles": water-like and oil-like, where a surfactant molecule is
composed of a few water-like particles (the head group) and a chain of
oil-like particles (the tail). The particles interact via a truncated
Lennard-Jones potential,
\begin{equation}
  V(r) = \left\{\begin{array}{ll}
  4\epsilon \left[ \left(\frac{d}{r}\right)^{12} -
  \left(\frac{d}{r}\right)^6 \right] &~~~ r\leq r_\rmc\\
  0 & ~~~r>r_\rmc,
  \end{array}\right.
\end{equation}
where $r$ is the inter-particle distance, $\epsilon$ the energy
parameter of the Lennard-Jones potential, $d$ its length parameter,
and $r_\rmc$ a cutoff. This potential has a minimum at
$r_{\rm min}=2^{1/6}d$.
Hence, for $r_\rmc\leq r_{\rm min}$ the potential is purely
repulsive, which is what is chosen for the oil--water interaction. For
$r_\rmc > r_{\rm min}$ the potential contains a short-ranged repulsion
followed by an attractive region, which is a suitable choice for the
water--water and oil--oil interactions. In addition, the particles
constituting a single surfactant molecule are connected by harmonic
potentials of equilibrium length $d$ and strong spring constant (much
larger than $\epsilon/d^2$), ensuring chain connectivity.

The MD simulation starts from a random distribution of surfactants in
water. It then evolves in time according to the classical equations of
motion governing the motion of individual particles.  The simulations
typically contain few ten thousands particles and are run for about
$10^5$--$10^6$ time steps \cite{Smit90,Smit93,Karaborni,Maiti02}.
Thanks to the coarse-grained description, this can amount to about 1
$\mu$s in real system time \cite{Maiti02}. Such a scheme was shown to
successfully reproduce the structure of monolayers and micelles of
various shapes, and to provide some understanding of the dynamics of
surfactant self-assembly \cite{Smit90,Smit93}.  On the other hand, as
in the MC case, the coarse-grained representation prevents a well
defined correspondence between the simulated system and the actual
molecules in the experiments.

\subsection{Phase behavior}
\label{sec_phase_reg}

Concentrated surfactant solutions and ternary water--oil--surfactant
systems exhibit a rich variety of disordered and liquid-crystalline
phases \cite{Schick,Benshaul,Safran}. Some examples are the lamellar
($L_\alpha$) phase, sponge ($L_3$) phase, hexagonal ($H_1$) phase, and
cubic ($V_1$) phase.  All of these phases are based on various packing
of surfactant layers --- bilayers (in the binary-mixture case) or
monolayers (in the ternary case). The lamellar phase is made of stacks
of parallel layers, the sponge phase contains a disordered arrangement
of multiconnected layers, the hexagonal phase consists of hexagonal
arrays of parallel cylinders, and in the cubic phase the surfactant
layers are spheres arranged in a cubic lattice.

Unlike micellization one deals here with macroscopic bulk phases and
their corresponding phase transitions. The powerful tools of
thermodynamics and statistical mechanics are hence
applicable. Consequently, the theory of surfactant phase behavior has
reached a more advanced level, in particular, in the case of phases
with long-ranged order. We shall not review these theories here, as
most of them have not been used in current models of gemini
surfactants, but merely mention the various approaches.

Two phenomenological approaches to the phase behavior of surfactant
binary and ternary mixtures have been used. The first is based on the
Ginzburg--Landau formalism, which has been widely used in statistical
physics \cite{Schick}. It starts with a lattice description of the
mixture and derives from it a coarse-grained, continuous expression
for the energy (Hamiltonian), which can be studied by
statistical-mechanical techniques. The second approach is based on the
elastic and thermodynamic properties of the membranes that make the
various phases. For a review see, \eg Ref.\ \cite{Safran}.  In
addition to these phenomenological theories, a variety of lattice spin
models employing Monte Carlo simulations, as discussed in Sec.\
\ref{sec_micelle_reg}, were originally designed and applied to study
surfactant phase behavior
\cite{Widom,Larson1,Larson2,Stauffer93,Stauffer94a,Stauffer94b}.

\section{Models of gemini surfactants}
\label{sec_gemini}

Having provided the necessary background, we now turn to models of
gemini surfactants. As will be demonstrated below, these models are
essentially extensions of surfactant self-assembly theories,
which have been reviewed in Sec.\ \ref{sec_regular}.  The binding
of the surfactant molecules into pairs via spacer chains introduces
new constraints affecting the molecular arrangement in monolayers,
micelles and mesophases, as well as the thermodynamics of self-assembly.

\subsection{Surface behavior}
\label{sec_surface_gemini}

\begin{figure}[tbh]
\vspace{.5cm}
\centerline{\resizebox{0.55\textwidth}{!}
{\includegraphics{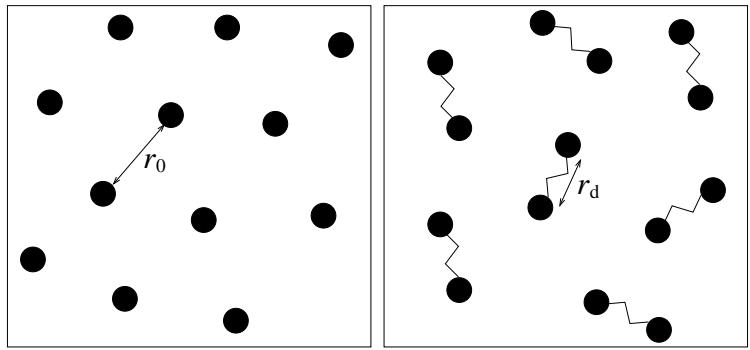}}}
\caption[]
{Schematic top view of a saturated monolayer at the water--air
interface.  Left: regular surfactant head groups separated by a mean
optimal distance $r_0$.  Right: Same view of gemini surfactants where
the head groups are linked into dimers by spacers.  The mean distance
between head groups in a dimer is $r_\rmd$, which in general differs
from the distance between unlinked head groups.}
\label{fig_surface_scheme}
\end{figure}

As in Sec.\ \ref{sec_regular}, we begin by looking at a saturated
monolayer, this time made of gemini surfactants, lying at the
water--air interface.  The gemini nature of the molecules, \ie the
introduction of the spacer, adds considerable complexity to the
problem, mainly since it introduces anisotropy and inhomogeneity into
the monolayer. A schematic view of the surface covered with dimers
is shown in Fig.\ \ref{fig_surface_scheme}. The dimers may be
oriented in various directions, and the distances between two linked
monomers (in a dimer) and between two unlinked ones will differ in general.

Nevertheless, we are going to disregard these complications and focus
on the simplest question: how does the introduction of a spacer
consisting of $s$ groups affect the inplane distance $r_\rmd(s)$
between two monomers belonging to the same dimer? Although seemingly
oversimplified, the answer to this question will give us key insight
into the surface behavior of gemini surfactants.

We proceed by reviewing and slightly extending the work presented in
Refs.\ \cite{us1,us2}.  We have seen in Sec.\ \ref{sec_surface_reg}
that the interaction between surfactants in a saturated
monolayer can be roughly approximated by ``effective springs'' whose
equilibrium length $r_0$ is determined by the optimum packing at
saturation, and whose spring constant $k_0$ is given in Eq.\
(\ref{ki}). It is thus natural to consider the spacer chain as another
``effective spring'', of equilibrium length $r_\rms$ and spring
constant $k_\rms$, connecting the two surfactant heads in a dimer. The
combination of the two types of monomer--monomer interaction --- the
one present between unlinked monomers and the one due to the spacer
--- is then reduced to adding together two springs in parallel. From
this we obtain
\begin{equation}
  r_\rmd(s) = \frac{k_0r_0 + k_\rms(s) r_\rms(s)}{k_0 + k_\rms(s)}.
\label{rd}
\end{equation}

The origin of the spacer ``spring'' is entropy, and its parameters are
determined by the statistical distribution of spacer configurations.
The ``equilibrium length'' of the spring is the mean end-to-end
distance of the spacer chain, and the ``spring constant'' is inversely
proportional to the variance of the end-to-end distance,
\begin{equation}
  r_\rms = \langle r\rangle,\ \ \ \
  k_\rms = \frac{\kT}{\langle r^2\rangle - \langle r\rangle^2},
\label{moments}
\end{equation}
where the averages are taken over all spacer chain configurations.
Thus, the harmonic-spring approximation for the spacer is equivalent
to representing the actual statistical distribution of spacer
configurations by its first two moments.

Before considering specific models for the spacer chain, let us
examine what qualitative results are expected from this description.
When the spacer is very short and rigid, such that $k_\rms \gg k_0$,
the equilibrium length $r_\rmd$ of the dimer is determined by the
spacer, $r_\rmd\simeq r_\rms$. On the other hand, when the spacer is
very long and flexible, such that $k_\rms \ll k_0$, $r_\rmd$ will be
determined by the regular monomer--monomer interaction, $r_\rmd\simeq
r_0$. Hence, upon increasing the number $s$ of groups in the spacer, we
expect $r_\rmd(s)$ to first increase and then saturate toward $r_0$, the
optimal distance between the monomeric surfactants. Whether the
behavior for intermediate spacer lengths is monotonous or not depends
on specific details of the spacer chain. If the spacer stiffness
$k_\rms(s)$ drops sufficiently fast with $s$, the ``interaction
spring'' will start dominating before $r_\rms(s)$ exceeds $r_0$, and
$r_\rmd(s)$ will then grow monotonously with $s$. By contrast, if
$k_\rms(s)$ decreases slowly with $s$, the ``spacer spring'' will
dominate even for quite long spacers and $r_\rmd\simeq r_\rms$ will
become larger than $r_0$. For even longer $s$, it will have to {\it
decrease} back toward $r_0$, leading to non-monotonous behavior in
this case.

\begin{figure}[tbh]
\vspace{.5cm}
\centerline{\resizebox{0.5\textwidth}{!}
{\includegraphics{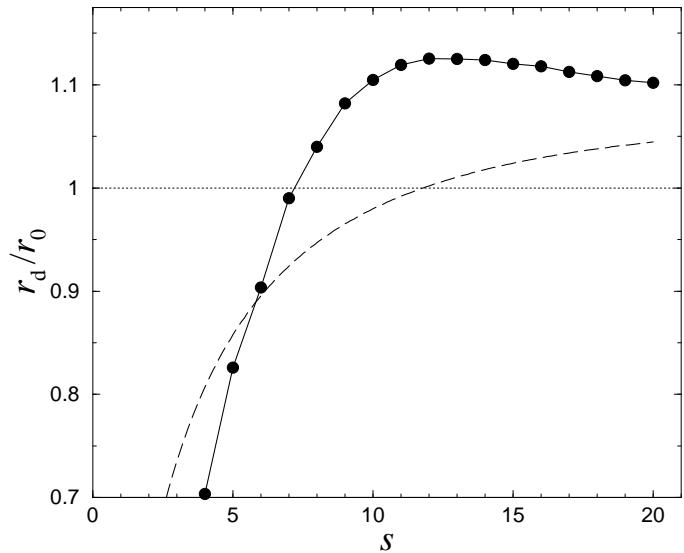}}} 
\caption[]
{The distance $r_\rmd$ between the two head groups of a gemini
surfactant in a saturated monolayer as a function of the number of
groups $s$ in its spacer. The distance is rescaled by $r_0$, the
distance between unlinked heads in a saturated monolayer of the
corresponding monomeric surfactant. Curves for two spacer models are
shown: Gaussian chain with no constraints (dashed line) and
rotational-isomeric chain restricted to the non-aqueous side of the
interface (symbols and solid line). The latter has a maximum at
$s=12$, in accord with experiments \cite{Alami,Perez}.}
\label{fig_surface}
\end{figure}

The simplest model for the spacer is that of a Gaussian,
constraint-free chain. This case is  somewhat artificial and is
discussed here merely as a model for very flexible and long chains,
in contrast with more realistic models discussed below for more rigid
chains.
A Gaussian chain consisting of $s$ segments is analogous to a random
walk of $s$ steps. The mean-squared displacement of such a walk,
averaged over all $s$-step walks, should scale linearly with $s$. The
mean end-to-end distance of a Gaussian spacer is therefore $r_\rms\sim
bs^{1/2}$, where $b$ is the segment length. More specifically, the
statistical distribution of the end-to-end distance in a Gaussian
chain is
\begin{equation}
  p(r)\rmd r = \left(\frac{3}{2\pi sb^2}\right)^{3/2}
  \rme^{-3r^2/(2sb^2)} 4\pi r^2 \rmd r.
\end{equation}
From the mean and variance of this distribution we get, according
to Eq.\ (\ref{moments}),
\begin{equation}
  r_\rms(s) = b \left(\frac{8s}{3\pi}\right)^{1/2},\ \ \
  k_\rms(s) = \frac{\kT}{[1-8/(3\pi)]b^2 s}.
\label{rnkn_gauss}
\end{equation}
Thus, in order to calculate $r_\rms$ and $k_\rms$ we just need to know
the segment length $b$. For a polymethylene spacer $b$ is 2.53\,\AA.
The remaining information required to compute $r_\rmd$ from Eq.\
(\ref{moments}) are the properties of the {\it monomeric} surfactant
in a saturated monolayer, namely $k_0$ and $r_0$.  In a roughly
hexagonal arrangement of molecules one has $k_0\simeq 0.3$
$\kT$/\AA$^2$ (see Sec.\ \ref{sec_surface_reg}). A saturated monolayer
of DTAB surfactants, for example, is known to have $a_0\simeq
55$\,\AA$^2$, \ie $r_0\simeq 8$\,\AA. By using these values and
substituting Eq.\ (\ref{rnkn_gauss}) in Eq.\ (\ref{rd}), $r_\rmd$ as
function of $s$ is calculated and depicted as the dashed line in Fig.\
\ref{fig_surface}.  The inter-monomer distance increases moderately
with $s$ and even exceeds $r_0$, yet the maximum and descent back to
$r_0$ are shallow and occur at very large $s$, lying outside the
experimentally relevant range of spacer lengths, $1\leq s \leq 20$
shown on the figure.

More realistically, the spacer chain can be described by the
rotational-isomeric model, where each segment in the chain can have
only three possible orientations with respect to the two that precede
it in the sequence (the three conformations called {\it trans} and
{\it gauche}$^\pm$) \cite{Flory}. In addition, we require that the
hydrophobic spacer be restricted to reside in the non-aqueous side of
the water--air interface. These constraints stiffen the chain and bias
its statistics toward larger end-to-end distances. We therefore expect
a larger overshoot and a sharper maximum of $r_\rmd(s)$, as is
confirmed by the solid curve in Fig.\ \ref{fig_surface}. The points of
this curve were obtained from simulations of rotational-isomeric
polymethylene chains whose ends were fixed to a surface (the
air--water interface) and whose segments were forbidden from crossing
that surface into the water side (see Ref.\ \cite{us1} for more
details). From the simulations one obtains the end-to-end distance
distribution and then extracts the ``spring'' parameters $r_\rms$ and
$k_\rms$ according to Eq.\ (\ref{moments}) \cite{us2}.

What has been calculated is the inter-monomer distance in a dimer and
not the average area per dimer in the monolayer. However, because the
latter must increase together with the former (cf.\ Fig.\
\ref{fig_surface_scheme}), this very simple ``spring'' model
reproduces the experimental observation of a non-monotonous behavior
of $a(s)$ for the \msm gemini surfactants \cite{Alami} and gemini
surfactants derived from arginine \cite{Perez}. While the shape of the
experimental curve is reproduced only qualitatively, the position of
the maximum at $s\simeq 12$ is the same as the one found
experimentally for 12-$s$-12 surfactants \cite{Alami}. Furthermore,
the spring model elucidates the source of the non-monotonous behavior
--- a competition between the regular monomer--monomer interactions on
one side, and the natural length and rigidity of the spacer, on the
other. According to this picture we should expect a more moderate and
monotonous increase in $a$ for more flexible spacer chain, as has been
demonstrated by the Gaussian-chain example above. This may explain the
behavior of $a(s)$ observed for the \meom gemini surfactants, having
more flexible poly(ethylene oxide) spacers
\cite{Dreja99,Zhu}. (Compare, for example, our Fig.\ \ref{fig_surface}
with Fig.\ 1 of Chapter 4 in this volume.)

These qualitative features, as well as the maximum at $s\simeq$ 10--12,
were found to remain unchanged upon various refinements of the
model, \eg the inclusion of non-bonded interactions within the spacer
chain, or a more detailed treatment of the monolayer structure
\cite{us1}.  A hydrophobic effect, \ie repulsion of spacer monomers
from the water phase, was invoked in several works as an explanation
for a ``lift-off'' of the spacer from the water surface and hence the
maximum in $a(s)$.  Such an effect, according to the spring
model, actually suppresses the maximum as it brings the spacer ends
closer together and thus reduces the overshoot of $r_\rmd$. We note
that this effect might be related, though, to the maximum observed in the cmc
of \msm surfactants at lower $s$ values; see Sec.\
\ref{sec_micelle_gemini}.

The description provided by the spring model is too simplistic to
account for various details of gemini surfactant monolayers. In
particular, two critical comments can be made. First, the
experimentally observed decrease of $a(s)$ for $s>12$ is much
steeper than what the model describes \cite{Alami}. As has been
suggested in Ref.\ \cite{Zana_review}, this might be a result of
increased pre-micellar aggregation in the bulk solution as the spacer
becomes more hydrophobic. Second, the model regards the spacer as an
isolated chain, whereas in reality the gemini surfactant has two
additional tail chains nearby. In this respect the model treats the
geminis as equivalent to ``bolaform'' surfactants. While undoubtedly
the presence of the tails is important for quantitative predictions,
it is not expected to alter the qualitative competition picture
described above, and a similar non-monotonous behavior of $a(s)$
was indeed observed in bolaform surfactants as well \cite{Menger}.

\subsection{Micelles}
\label{sec_micelle_gemini}

The micellization behavior of gemini surfactants is qualitatively
different from that of regular ones. We have reviewed some of these
differences in the Introduction, and they can now be further
elucidated in the light of what we have discussed in the previous
sections.

The cmc of gemini surfactants is typically one to two orders of
magnitude lower than that of the corresponding monomeric surfactants
\cite{Zana91}. The lower cmc can be directly attributed to the
increase in the number of hydrocarbon groups in the molecule, \ie
decrease in the hydrophobic contribution $g_{\rm hc}$, due to the
second tail, and also due to the hydrophobic spacer
chain in the case of \msm surfactants.  Based only on the contribution of a
second tail to $g_{\rm hc}$ and the fact that the molecular volume is
roughly doubled going from the monomeric surfactant to a gemini one,
one would have predicted a {\it larger} decrease in cmc than what is
actually observed. The difference is probably due to unfavorable terms
introduced by the spacer, which will be further discussed below.

The cmc of \msm gemini surfactants, instead of monotonously decreasing
with the number $s$ of spacer hydrocarbon groups (\ie with molecular
hydrophobicity), is a non-monotonous function with a maximum around
$s\simeq$ 4--6 \cite{Devinsky,Zana91,De}.
A corresponding non-monotonous behavior is observed in the
Krafft temperature \cite{Zana02} and micellization enthalpy
\cite{Grosmaire}. This behavior can be attributed to the straightness and
rigidity of short spacers, which force their hydrocarbon groups to be
in unfavorable contact with water. At about $s\simeq$ 4--6, although
the spacer chain is still rigid, a {\it gauche} conformation should
become accessible, allowing some of the groups to penetrate in the
micellar hydrophobic core.  When the spacer is hydrophilic this effect
should be absent, as indeed is the case with
\meom surfactants, exhibiting a weak monotonous increase
of the cmc with the hydrophilic spacer length, $x$
\cite{Dreja99}.

As a function of spacer length $s$, \msm surfactants exhibit an
unusual progression of aggregate shapes from cylinders to spheres to
bilayers. This is different from the more natural succession,
occurring in monomeric surfactants, where the change in aggregate
curvature is monotonous: spheres transforming into cylinders
transforming into bilayers.  Assuming that the molecular area at the
aggregate surface is related to that in a saturated monolayer, this
uncommon behavior can be qualitatively understood in view of the
non-monotonous variation of $a(s)$ as function of $s$, discussed
in Sec.\ \ref{sec_surface_gemini}.  Considering that the radius and
volume of the micellar core depend primarily on the tails and not on
the spacers, an increase and then a decrease of $a$ as a function
of $s$ should be accompanied by a decrease and then an increase in the
packing parameter $P$ of Eq.\ (\ref{packing}), hence the unusual
morphological sequence.

More specific predictions require a detailed theory and will be reviewed
next.

\subsubsection{Phenomenological model}

An extension of the phenomenological theory of surfactant aggregation
to gemini surfactants with hydrophobic spacers is presented in Ref.\
\cite{Nagarajan2}. It introduces the following additions and
modifications to the model outlined in Sec.\
\ref{sec_micelle_reg}.

(i) The hydrophobic free energy $g_{\rm hc}$ contains, apart from the
double-tail contribution, also a spacer contribution. Only the spacer
section which penetrates in the micellar hydrophobic core, $s_{\rm
core}$, is considered.  This spacer section is taken simply as the
difference between the total spacer length and the mean head--head
distance, $s_{\rm core}=s-a^{1/2}/b$. Since the three chains (two
tails and spacer) are in partial contact already prior to aggregation,
the hydrophobic energy gain per hydrocarbon group is taken to be
smaller than in the case of a single-tail surfactant.

(ii) The interfacial term, $g_{\rm int}(a)$, is modified to account for
the part of the core--water interfacial area that is now occupied by
the spacer.  The chain length that participates in this ``shielding''
is proportional to $a^{1/2}$. This contribution is thus
\begin{equation}
  \delta g_{\rm int} \simeq (\gamma_2-\gamma_1)a^{1/2}w,
\end{equation}
where $\gamma_2$ is the spacer--water interfacial tension and $w$ is
the spacer width. If the spacer is a polymethylene chain then
$\gamma_2=\gamma_1$ and this correction vanishes.

(iii) When the spacer is short, it forces the two tails to be closer
together than they would be if they belonged to two separate
molecules. This packing constraint reduces the entropy of the tail
chains.  For a single-tail surfactant, the area close to the
core--water surface sampled by tail groups is $a_{\rm tail}\sim
v/R$ (cf.\ Fig.\ \ref{fig_packing_parameter}), with a prefactor of
order unity that varies with aggregate shape
\cite{Nagarajan1,Nagarajan2}. The proximity to a second tail due to the
spacer reduces this available area per tail to $a_{\rm sp}\sim
(sb)^2$. Thus, the contribution to the free energy can be estimated as
\begin{equation}
  g_{\rm tail} \simeq \kT \ln \left(
  \frac{a_{\rm tail}}{a_{\rm sp}} \right) \simeq
  \kT \ln \left( \frac{v}{Rb^2s^2} \right).
\end{equation}

(iv) Finally, the most difficult modification to handle is the
electrostatic one. A short spacer forces the distance between two
connected head groups to be shorter than that between two unconnected
ones, resulting in a nonuniform charge distribution of pairs
over the micellar surface (cf.\ Fig.\ \ref{fig_surface_scheme}).
This problem is bypassed in Ref.\
\cite{Nagarajan2} by introducing an empirical correction factor to
$g_{\rm es}$, which becomes equal to unity when the spacer is longer
than the mean inter-head distance.

This extended phenomenological model is applied in Ref.\
\cite{Nagarajan2} to gemini surfactants with short hydrophobic
spacers, using parameters known from regular single-tail and double-tail
surfactants. The model yields cmc values for various tail
lengths in good agreement with the measured ones. More
important, it correctly accounts for the observed micelle shapes of
\msm surfactants with small $s$, \ie the crossover from cylinders to
spheres as $s$ is increased. Following the changes in the various free
energy contributions, one can identify the crossover mechanism as a
competition between the elastic and packing contributions from the
tails (favoring cylinders), and the electrostatic contribution
(favoring spheres). Note, however, the various assumptions and
approximations involved in these calculations. Although the
electrostatic contribution to the free energy is found
to be crucial for the self-assembly behavior,
it is treated somewhat dubiously, as already admitted in Ref.\
\cite{Nagarajan2}.

\subsubsection{Computer simulations}

The additional complexity introduced by the spacers makes analytical
calculations very difficult. One is inclined, therefore, to resort to
computer simulations in order to gain detailed information on the
self-assembly of gemini surfactants.

\begin{figure}[tbh]
\vspace{.5cm}
\centerline{\resizebox{0.35\textwidth}{!}
{\includegraphics{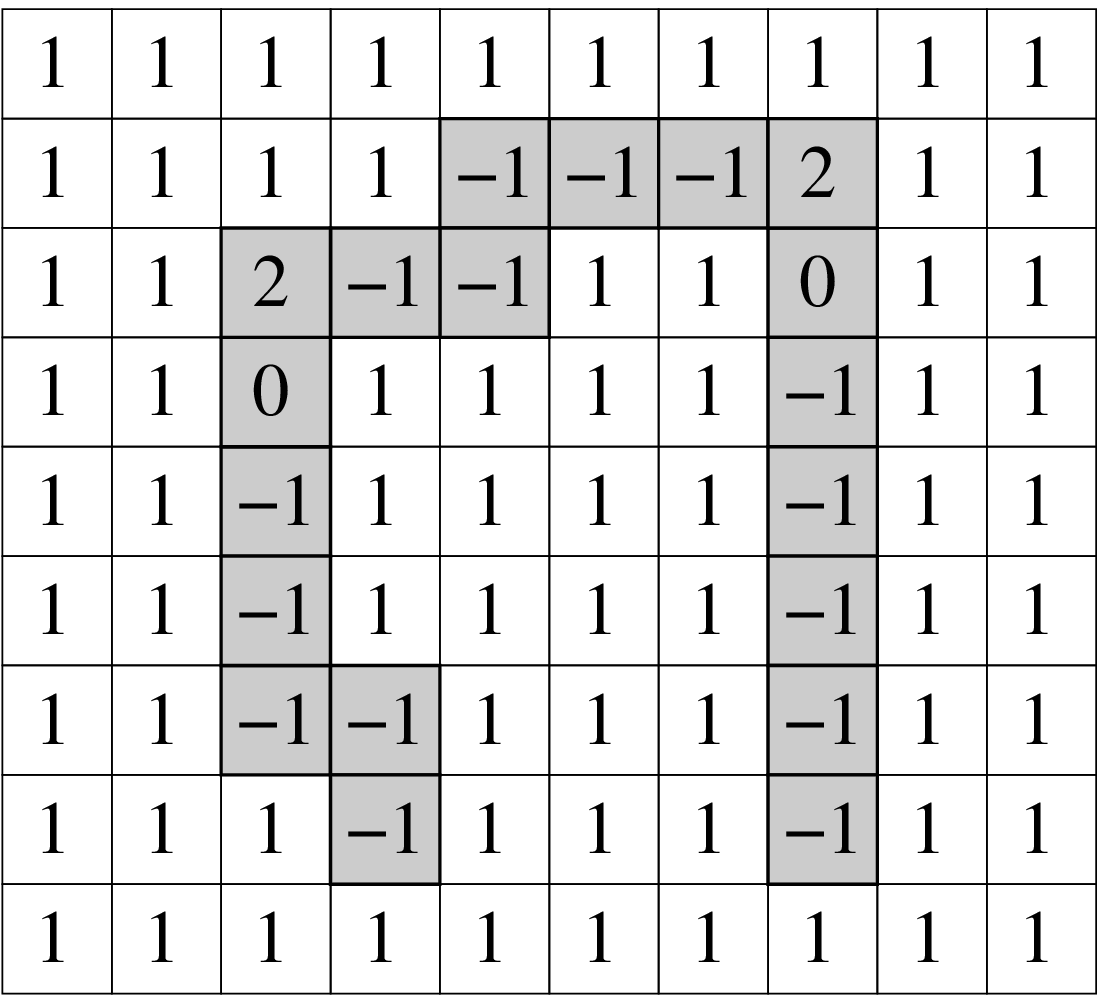}}}
\caption[]
{Schematic representation of a gemini surfactant molecule (gray)
solubilized in water (white) in a lattice spin model
\cite{Maiti1,Maiti2}. The spin scheme is similar to that of Fig.\
\ref{fig_Stauffer}, except that the two hydrophilic head groups (spin $+2$)
of each surfactant are linked by a spacer chain. The spacer is
composed of spins $-1$ for hydrophobic spacers, as shown here, or $+1$
for hydrophilic spacers. In addition, a ``kink'' in the spacer chain
such as the one shown here is assigned an energy penalty in order to
mimic the role of spacer stiffness.}
\label{fig_Maiti}
\end{figure}

The statistical-mechanical approach based on Monte Carlo (MC) simulations,
as outlined in Sec.\ \ref{sec_micelle_reg}, was extended to treat
gemini surfactants \cite{Maiti1,Maiti2}. The ``spin'' assignment to
various groups and the corresponding energy function are the same
as for regular, monomeric surfactants [see Eq.\ (\ref{Ising})]. The main
modification is the connection of head groups in pairs via spacers
(Fig.\ \ref{fig_Maiti}). Both hydrophobic spacers (spins $\sigma=-1$)
and hydrophilic ones ($\sigma=+1$) were simulated. In addition, the
role of spacer stiffness was checked by assigning an energy
penalty for ``kinks'' in the spacer configuration.

\begin{figure}[tbh]
\vspace{.5cm}
\centerline{\resizebox{0.4\textwidth}{!}
{\includegraphics{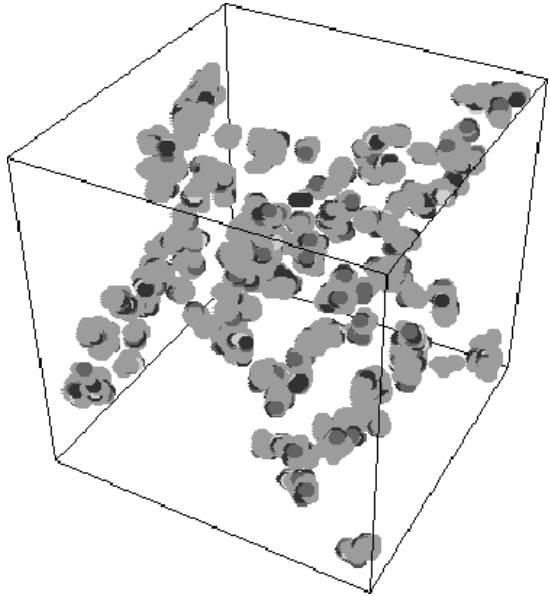}}}
\caption[]
{Worm-like micelles formed in a Monte Carlo simulation of the spin
model. The gemini surfactant has two heads of one lattice site each,
two neutral groups of one site each, two tails of 15 sites each, and a
hydrophobic spacer of two sites. Different gray tones correspond to
different surfactant groups. The water molecules are not shown for
clarity.  (Reprinted with permission from Ref.\ \cite{Maiti1}.)}
\label{fig_Maiti_worm}
\end{figure}

This spin model reproduces a few important properties of gemini
surfactants as observed in experiments, primarily the non-monotonous
dependence of the cmc on spacer length for hydrophobic spacers, and
the formation of branched and entangled worm-like micelles in the
case of short hydrophobic spacers (see Fig.\
\ref{fig_Maiti_worm}). However, the MC simulations produce also some
findings which are not in full accord with experiments.  The cmc is
found to {\it increase} with tail length, unlike the common experimental
results.  (An exception to this
rule is presented in the experiment of Ref.\
\cite{Menger93}, yet in a system having equilibration problems.)
The mechanism for such a cmc increase with
surfactant hydrophobicity is unclear. It is hard to simply attribute it
to spacer--head repulsion, since the increase is found to be
insensitive to the ``spin'' associated with the head group.
A similar issue appears in
the cmc dependence for hydrophilic spacers, which is found to {\it
decrease} with spacer length, in disagreement with the experimentally
observed (and
expected) increase \cite{Dreja99}. The maximum in the cmc as a function
of $s$ is obtained for long hydrophobic spacers of about
$s\simeq 12$ regardless of tail length, contrary to the experimental
result of only $s\simeq 5$ \cite{Devinsky,Zana91,De}.  Surfactants
with long ($s=16$) hydrophobic spacers are found to form rod-like
cylindrical micelles, whereas in experiments they form bilayers
\cite{Danino}.
Hence, the spin model investigated by MC simulations seems
to capture part of the essential features
of gemini surfactant self-assembly while missing others.

\begin{figure}[tbh]
\vspace{.5cm}
\centerline{\resizebox{0.35\textwidth}{!}
{\includegraphics{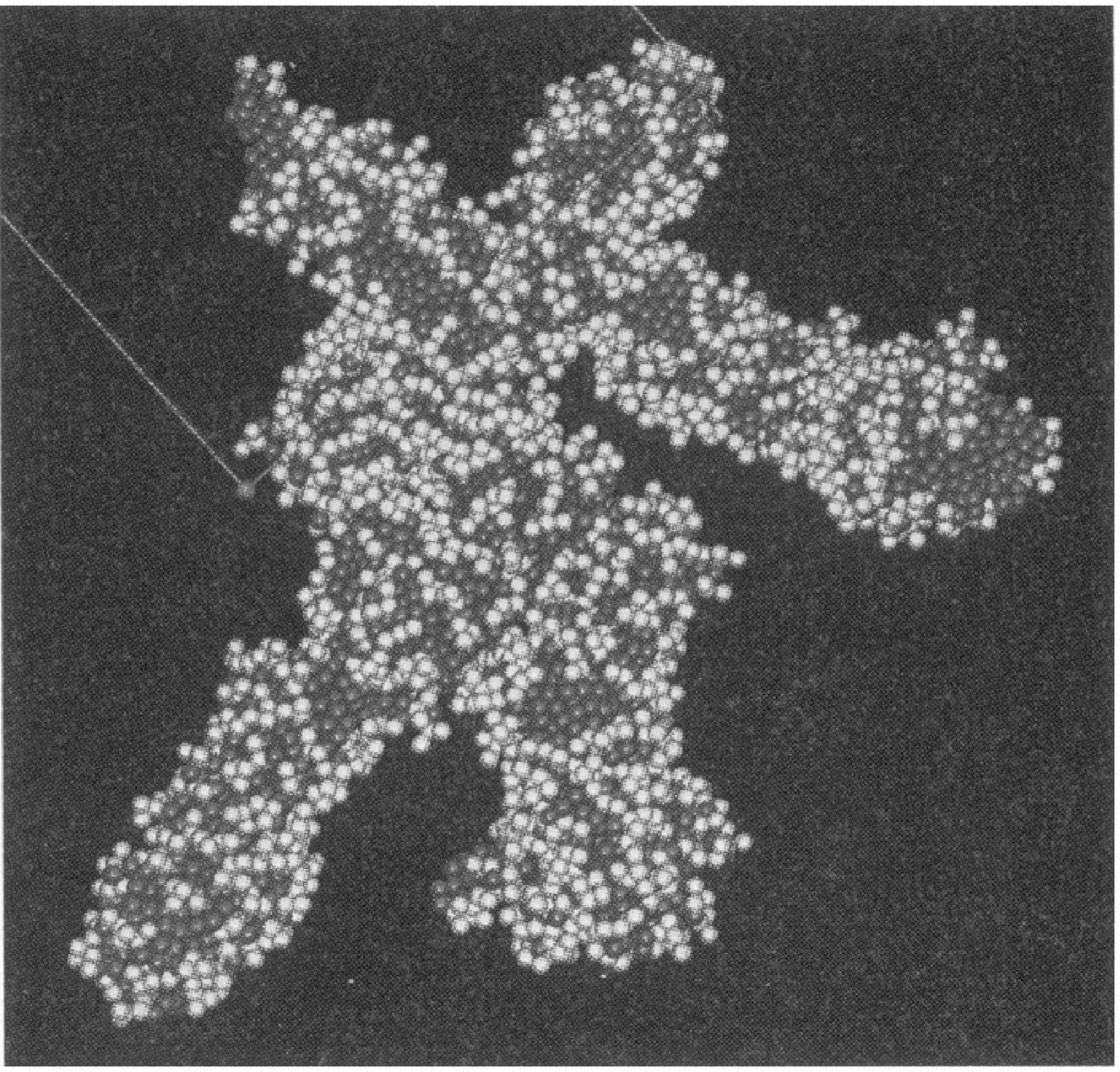}}
\hspace{.5cm}
\resizebox{0.35\textwidth}{!}
{\includegraphics{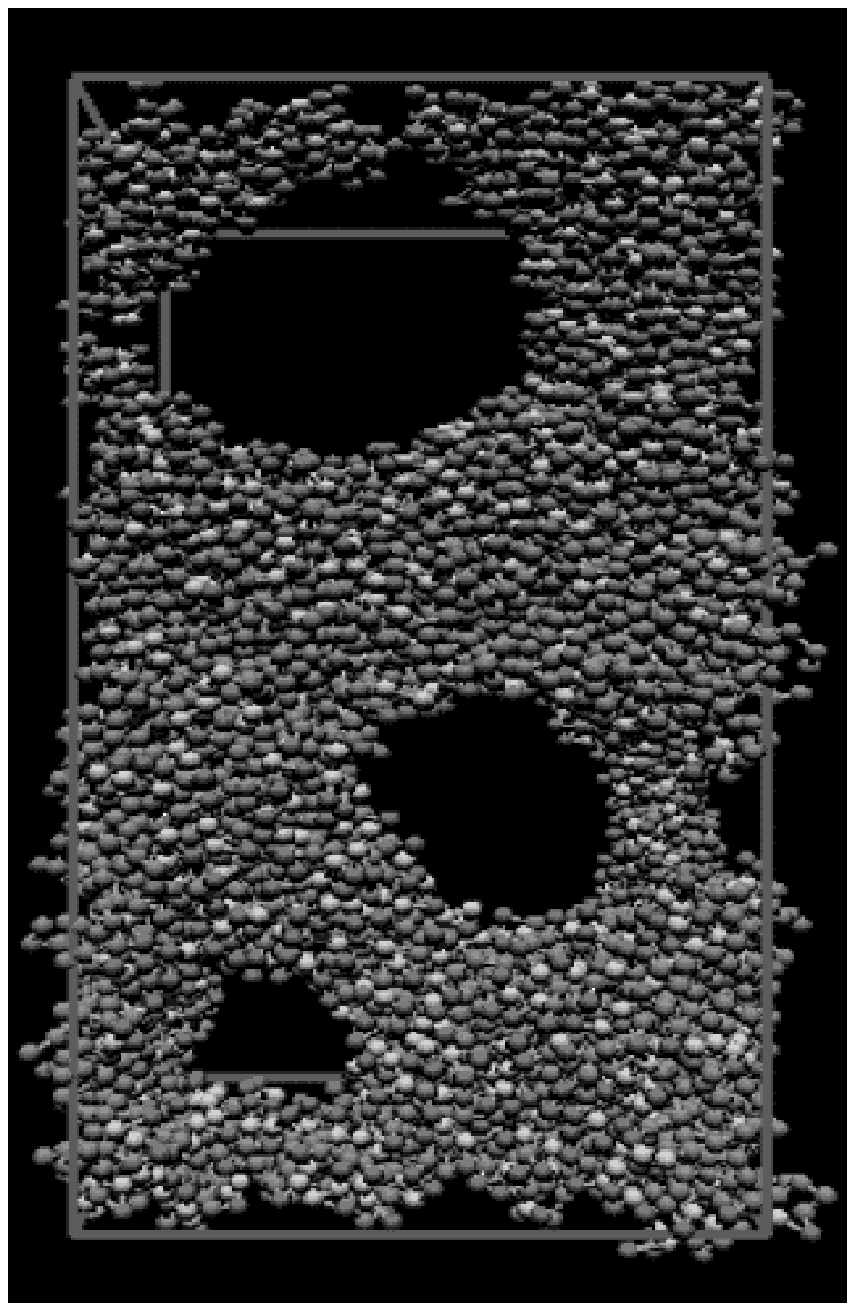}}}
\caption[]
{Micelles formed in coarse-grained Molecular Dynamics
simulations of gemini surfactants.  
Left: branched worm-like micelle. The surfactant has two heads of
three water-like particles each, two tails of six oil-like particles
each, and a spacer of one oil-like particle. Different gray tones
correspond to different surfactant groups. The water molecules are not
shown for clarity.  (Reprinted with permission from Ref.\
\cite{Karaborni}.  Copyright 1994 American Association for the
Advancement of Science.)
Right: ring micelles.  The surfactant has two heads of one water-like
particle each, two tails of four oil-like particles, and a spacer of
two oil-like particles. Different gray tones correspond to different
surfactant groups. The water molecules are not shown for clarity.
(Reprinted with permission from Ref.\
\cite{Maiti02}.  Copyright 2002 American Chemical Society.)
}
\label{fig_MD}
\end{figure}

The ``bead--spring'' Molecular Dynamics (MD) approach discussed in Sec.\
\ref{sec_micelle_reg} was extended as well to treat gemini surfactants
\cite{Karaborni,Maiti02}. The only essential modification is the
connection of head groups in pairs by spacer chains. Like the tail
chains, the spacers are made of ``oil-like'' particles connected to one
another by harmonic springs, where only hydrophobic spacers were studied.
These MD simulations are able to
reproduce the micellar shapes formed by the \msm gemini surfactants ---
branched worm-like micelles and ring micelles ---
compared to the spherical morphology formed by the corresponding
monomeric surfactants (see Fig.\ \ref{fig_MD}).
A similar coarse-grained MD approach, along with a
self-consistent-field calculation, were applied to the more complex
glucitol amine gemini surfactants, which have flexible sugar
side-chains attached to the charged head groups \cite{vanEijk}.
The main finding is a transition from cylindrical micelles to
bilayers upon increasing $p$H, in accord with experimental
indications \cite{Fielden}.

\subsection{Phase behavior}
\label{sec_phase_gemini}

The spacer length has an unusual effect also on the phase behavior of
systems containing \msm gemini surfactants. This fact has been
mentioned already in the Introduction. For binary surfactant--water
mixtures, the regions in the concentration--temperature phase
diagrams, where single-phase hexagonal and lamellar phases are the
stable state, become smaller as $s$ increases, vanish for
$s=10$--$12$, and then are finite again for $s=16$ \cite{Zana93b}. In
ternary water--oil--surfactant phase diagrams of the same surfactants,
the size of the emulsification (single-phase) region increases with
$s$ and then decreases, with a maximum at $s\simeq 10$
\cite{Dreja98a}.

These observations can be rationalized in the light of the
packing considerations discussed in Secs.\
\ref{sec_surface_gemini} and \ref{sec_micelle_gemini} \cite{Zana_review}.
As $s$ increases from small values, the optimal area per molecule
$a(s)$ increases and the packing parameter $P$ of Eq.\
(\ref{packing}) decreases. Hence, surfactant packing into bilayers,
which are the building blocks of the lamellar and hexagonal
mesophases, becomes less and less favorable. As we have seen in the
previous sections, this effect is maximal for $s=10$--$12$, whereupon
bilayers apparently can no longer be stabilized. When $s$ increases
further, $a$ decreases, and bilayers can form again. This
qualitative description agrees also with the experimental results for
\meom gemini surfactants. The observed monotonous increase in the
phase-diagram region belonging to the isotropic micellar phase
\cite{Dreja98b} is in accord with the moderate monotonous increase of
$a(s)$ observed for these surfactants at the water--air interface
\cite{Dreja99}.

In ternary oil--water--surfactant mixtures, as $s$ increases and 
$P$ decreases, the
surfactant monolayers required for stabilizing a microemulsion can
have higher curvature. Hence, smaller oil domains can form and the
microemulsion region in the phase diagram extends toward higher
surfactant concentration. This effect, too, should be maximal for
$s=10$--$12$, as observed in the experiment \cite{Dreja98a}.

A theoretical study of gemini surfactant phase behavior, using MC
simulations of a lattice model and a theory of mixture thermodynamics,
is presented in Ref.\ \cite{Layn}. Employing a simulation technique
which, after equilibration, samples the composition of small regions
in the entire lattice, the model is insensitive to long-ranged
structures and is rather focused on the thermodynamics of phase
coexistence. The study was restricted also to short hydrophilic
spacers.  Thus, the results cannot be compared with the experiments
mentioned above. The main finding is the suppression of the
three-phase region (coexistence of water-rich, oil-rich, and
surfactant-rich phases) upon introducing molecular rigidity.

\section{Conclusions and open questions}
\label{sec_conclusion}

The unusual self-assembly behavior of gemini surfactants poses
challenging puzzles to theoretical investigations. We have reviewed
the currently available models that attempt to address these puzzles,
concentrating on surface properties, micellization, and phase
behavior of gemini surfactant solutions. The overall impression
emerging from the current state of the art is that, despite several
successes, the theoretical understanding of gemini surfactants is
fragmentary and lags behind the wealth of available experimental
data.

As demonstrated in this review, current gemini surfactant models are
based on previous theories of surfactant self-assembly, with the most
essential modifications required due to the addition of the spacer
chains. It seems that this route has been exhausted, and further
progress will depend on detailed consideration of  features
distinguishing gemini surfactants from regular monomeric ones.

One of the distinct factors that stands out as a crucial ingredient is
the spacer effect on lateral organization of the surfactant molecules
in water--air monolayers and at aggregate surfaces. Linking the head
groups in pairs has at least three different aspects as can be 
seen in Fig.\ \ref{fig_surface_scheme}.
(i) The distribution of head groups on the surface becomes
inhomogeneous as linked head groups have a mutual distance
different from that of unlinked ones. This should affect, for example,
the surface charge distribution.
(ii) The spacers give the surfactant molecule an inplane orientation,
\ie the combined head group made of the two monomeric heads and spacer
is very anisotropic. Such breakdown of isotropy, as is known from
other systems, may lead to drastic effects
on the overall behavior and may result in the formation of inplane
liquid-crystalline order. Nematic ordering due to elongated head
groups was theoretically addressed in the case of bilayer membranes
\cite{Fournier}. A recent work combining dichroism spectroscopy and
atomistic MD simulations has revealed orientational ordering of gemini
surfactants in cylindrical micelles \cite{Oda}.  More theoretical work
is required to elucidate this issue.
(iii) The above two aspects apply as well to double-tail surfactants
(\eg phospholipids) with large, elongated head groups. What truly
distinguishes gemini surfactants from double-tailed surfactants is
the fact that the spacer makes a ``soft'' link between the head groups.
Containing at least several chemical bonds, it allows a
degree of conformational flexibility to the entire molecule, \eg with
respect to the relative orientations of the two tails. This feature is
probably what allows gemini surfactants to form such uncommon
structures as branched micelles and ring micelles. 
In the MD simulations of Ref.\ \cite{Karaborni}, for example, the gemini
surfactants residing in a branching junction of a worm-like micelle
were found to have their two tails oriented in different directions.
This property might also make gemini surfactants serve as cross-linkers of 
regular micelles \cite{Maiti00}.
Hence, it looks like our understanding of gemini surfactant
self-assembly will be incomplete until we have a good account of
the interplay between various lateral organizations of these molecules
at surfaces.

Another important direction where there is substantial experimental
information but almost no theory is the dynamics and rheology of
gemini surfactant solutions. This aspect is
particularly relevant to applications, as these solutions exhibit
unusual and useful rheological properties such as shear thickening at low
volume fractions \cite{Zana_review}. Moreover, recently these
properties have made micellar solutions of gemini surfactants a model
system for studying nonequilibrium behaviors such as shear thickening
and ultra-slow relaxation \cite{Oelschlaeger}. We note that the
dynamic issues and the issue of molecular organization mentioned above
may be closely related. It has been argued recently that the
distinct rheological behavior of worm-like micellar solutions (\eg
shear thickening) stems from the formation and interlinking of ring
micelles \cite{Cates}.

We hope that this review and the posed open questions will
motivate further theoretical studies of this class of
fascinating and very useful self-assembling molecules.


\acknowledgments

We are grateful to Raoul Zana for introducing us to the field of
gemini surfactants and for numerous discussions and suggestions.  We
also benefited from discussions with Igal Szleifer.  HD acknowledges
support from the Israeli Council of Higher Education (Alon Fellowship)
and the Yeshaya Horwitz Foundation (Complexity Science Fellowship).
DA would like to acknowledge partial support from the Israel Science
Foundation under grant no.\ 210/02, the Israel--US Binational Science
Foundation (BSF) under grant no.\ 98-00429, and the Alexander von
Humboldt Foundation for a research award.


\end{document}